\documentstyle[preprint,aps]{revtex}  
\begin{document}
\draft
\title{Quasiparticle properties of a coupled quantum wire 
electron-phonon system}
\author{E. H.\ Hwang$^1$, Ben Yu-Kuang Hu$^{1,2}$, and S.\ Das Sarma$^1$}
\address{$^1$ Department of Physics, University of Maryland, College Park, 
Maryland  20742-4111\\ 
$^2$ Mikroelektronik Centret, Danmarks Tekniske Universitet, Bygning 
345\o, DK-2800 Lyngby, Denmark}
\date{\today}
\maketitle
\begin{abstract}
We study leading-order many-body effects of longitudinal optical (LO)
phonons on electronic properties of one-dimensional quantum wire
systems. We calculate the quasiparticle properties of a weakly polar
one dimensional electron gas in the presence of both
electron-phonon and electron-electron
interactions. The leading-order dynamical screening approximation
(GW approximation) is used to obtain the electron self-energy, the
quasiparticle spectral function, and the quasiparticle damping rate in
our calculation by 
treating electrons and phonons on an equal footing.
Our theory includes effects (within the random phase approximation) of
Fermi statistics, Landau damping, plasmon-phonon mode coupling, phonon
renormalization, dynamical screening, and impurity scattering. In
general, electron-electron and electron-phonon many-body
renormalization effects are found to be nonmultiplicative and
nonadditive in our theoretical results for quasiparticle properties.
\end{abstract}

\newpage

\section{INTRODUCTION}

There has been a great deal of recent interest in
ultranarrow confined 
semiconductor systems, called quantum wire structures, where the
motion of the electrons is essentially restricted to be one
dimensional (1D) because of their potential, both, in
basic physics and for applications to novel device concepts\cite{one}.
Thus, both from the fundamental and applied physics viewpoints, there
is interest in understanding the electronic properties of
quasi-one-dimensional quantum wires.
The quantum wire structures are usually fabricated by lithographic
techniques\cite{non}, epitaxial growth on tilted
superlattices\cite{epit}, overgrowth of 
preetched V-groove patterns\cite{over}, stress patterning\cite{stre},
and cleaved edge overgrowth method\cite{clea} which has been used
to produce narrow quantum wires with atomic scale definition.

In a doped polar semiconductor (in which most of the quantum
wire structures are fabricated) free carriers couple to the
longitudinal optical (LO)-phonons of the underlying lattice through the
long-range polar Fr\"{o}hlich 
coupling. The carrier-LO phonon interaction leads to polaronic 
many-body renormalization of the single particle free carrier
properties, e.g., polaronic effective mass renormalization, lowering of
the effective band edge, inelastic scattering, broadening of the
quasiparticle spectral 
function, etc. Even though the weakly polar III-V materials have rather
small Fr\"{o}hlich coupling constants, the electronic properties can
still be substantially
modified by the Fr\"{o}hlich interaction, and, conversely, the
phonon properties are affected by the
Coulomb interaction between the electrons. Coulomb and Fr\"{o}hlich
interactions cannot be disentangled in experimental situations, because
many-body renormalization effects due to electron-electron and
electron-phonon interactions in such a
coupled system are fundamentally nonmultiplicative in nature (as well
as being nonadditive).

Our goal is to study a coupled 1D
electron-phonon many-body system treating electrons and phonons on an
equal footing. Direct interaction between electrons via the Coulomb
interaction and polar electron LO-phonon interaction (which is
fundamentally Coulombic in origin, arising from the dynamical
interaction between charge carriers and lattice ions) via the Fr\"{o}hlich
coupling are among the most extensively studied many-body
interactions in solid-state physics. In a two-dimensional electron
gas, quasiparticle properties of a polar electron gas have 
been extensively studied by treating the electron--electron and 
electron--phonon interactions on an equal footing\cite{ja}. 
In ref. \onlinecite{ja}, Jalabert and Das Sarma find that the Coulomb
and Fr\"{o}hlich interaction effects are nonmultiplicative, and the
actual many-body correction for a 2D
electron gas is substantially different from the one-polaron result 
(i.e. a single electron coupled to an LO-phonon system).
In spite of substantial current interest in the properties of a polar
quasi-one dimensional 
electron gas (1D EG) existing in GaAs quantum wires,
there has been no detailed quantitative study of 1D quasiparticle 
properties including both the
electron--electron and electron--phonon interactions.
In a GaAs (or, other III-V materials) quantum wire, 
the electronic energy scales (Fermi and plasmon energies) can be comparable
to the LO-phonon energy, which produces strong mode coupling between
plasmons and 
LO-phonon modes\cite{hwang}. Thus, a complete many-body analysis based
on treating 
electrons, phonons, and plasmons equivalently within the same
approximation scheme is needed. We provide such an analysis in this
article based on the leading-order many-body perturbation theory.
The Fr\"{o}hlich electron-LO phonon coupling constant in GaAs is small,
making a weak coupling diagrammatic expansion for calculating
quasiparticle properties a meaningful approximation.

Well-established theoretical results\cite{tl,ma}
(Tomonaga-Luttinger model) predict unusual non-Fermi liquid-like
properties for an 
interacting 1D system. No matter how weak the electron-electron
interaction the non-interacting Fermi surface is nonperturbatively
unstable in the presence of interaction.
While the instability of a noninteracting 1D Fermi gas to an
interacting 1D Tomonaga-Luttinger liquid is a well-established
theoretical concept, the actual quantitative effect associated with the
nonperturbative disappearance of the Fermi surface is, in fact, rather
small at finite temperatures in a weakly interacting 1D system such
as a GaAs quantum wire 1D EG. Most of the experimentally observed
electronic properties of semiconductor quantum wires have so far been
interpreted in the literature as effective 1D Fermi liquids, albeit at
finite temperature and in the presence of impurity scattering induced
collision broadening. Indeed, based on a weak-coupling diagrammatic
expansion in dynamically screened Coulomb interaction, it has recently
been argued\cite{hu} that 1D electrons in GaAs quantum wires behaves
for most practical purposes as a thermally broadened Fermi liquid at
finite temperatures and, most importantly, in the presence of impurity
scattering effects which are invariably present in real semiconductor
quantum wires. The physical reason behind this result is that impurity
scattering cuts off the emission of arbitrarily low energy plasmons by
overdamping them, and the instability of the Fermi surface, arising
from the decay of quasiparticles into low energy plasmons, is
prevented. Our approach in this paper is to include electron-LO phonon
Fr\"{o}hlich interaction in this theory by doing an expansion in
leading order dynamically screened Coulomb plus Fr\"{o}hlich
interactions, both in the presence and absence of impurity scattering
effects for the purpose of comparison. We note in this context that
(1) the leading order calculation\cite{hu}, in the absence of impurity
scattering, is consistent with the nonperturbative Tomonaga-Luttinger
result in the sense that it indicates a non-Fermi liquid type behavior
(with different analytic properties, however); (2) the
nonperturbative Tomonaga-Luttinger liquid type calculations have only
been carried out for the point contact model electron-electron
interaction whereas our work uses the true long range Coulomb (and
Fr\"{o}hlich) interaction as the bare interaction.

Usually the quantum wire confinement in the $z$-direction (taken to be
the growth direction) 
is much stronger than the confinement in the $y$ direction
in real systems (assuming the 2D EG to be confined in the $x-y$
plane). The electron is typically confined in the $z$-direction on
the order of less than $100\,\mathrm{\AA}$, whereas currently the best
confinement in the $y$ direction is approximately $300\,\mathrm{\AA}$, 
leading to
at least an order of magnitude difference in the energy-level spacings
of the $y$ (a few meV) and the $z$ (the order of tens of meV) direction.
In this paper we therefore assume for the sake of simplicity that the
electron gas has zero 
thickness in the $z$ direction, but has a finite width in the $y$
direction. This, in fact, is one of the common models used in studying
1D quantum wire structures. Confinement of the electrons in the
$y$ and $z$ 
directions leads to the quantization of energy levels into different
subbands. We assume the 1D quantum limit where the
energy separation between the lowest-energy (ground) and higher-energy
(excited) subbands
is so much larger than all other energy scales in the problem that the
higher subbands can be ignored.
In this paper we calculate in the 1D quantum limit the zero
temperature electron
self-energy from the effective total 
dynamical electron--electron interaction, which is calculated by
taking into account both the
Coulomb electron--electron interaction and the LO phonon-mediated
Fr\"{o}hlich interaction, within the random-phase approximation
(RPA). We treat electron--electron and electron--phonon interactions
equally, and our only approximations are the use of the
RPA and the neglect of vertex corrections, i.e. we keep the lowest
order diagram in the total effective dynamically screened
interaction. This approximation has long 
been employed successfully to calculate properties of three- and
two-dimensional electron systems\cite{hedin,ando}.
Furthermore, the RPA gives the {\em exact} low-energy screening 
properties of one-dimensional systems\cite{li1}. 
The Feynman diagrams for the self-energy and the effective
interaction in these approximations are shown in Fig. 1.
We ignore diagrams which are higher order in the screened interaction,
such as the diagram shown in Fig. 1(c).

This paper is organized as follows. In Sec. II we introduce our
model for the coupled 1D electron-LO phonon system. In Sec. III we
describe our self-energy and spectral function calculations for
the quantum wire at $T=0$. In Sec. IV we present our numerical results
for various quasiparticle properties. We provide a conclusion in Sec.
V. Throughout this paper we have used the terminology in which we
refer to the situation having both Coulomb 
and Fr\"{o}hlich interactions as the coupled system, whereas the
situation with only the Coulomb interaction (i.e. with the
Fr\"{o}hlich coupling turned off) is called the uncoupled
system. Also note that in this paper, $\hbar =1$.

\section{MODEL}

Our model consists of a 1D EG coupled to bulk
dispersionless LO-phonons at zero temperature.
Electrons interact among themselves through the Coulomb interaction
and through virtual-LO-phonon exchange via the Fr\"{o}hlich
interaction. In calculating the effective 1D electron-phonon
interaction we sum over the phonon wave vector in the other two
dimensions in the standard manner\cite{ja}. The Coulomb interaction is
logarithmically divergent 
in the 1D wave vector space, and therefore we use the more realistic
finite width quantum wire model\cite{li}. In our extreme quantum limit model
where only the lowest 1D subband is occupied by the electrons, we obtain the
Coulomb interaction matrix element $V_c(q)$ by taking the quantizing
confinement potential to be of infinite square well type\cite{hu,typo},
\begin{equation}
V_c(q)=\frac{2e^2}{\epsilon_{\infty}} \int^{1}_{0}dx\ K_0(qax) \left [ 
(1-x)\,(2+\cos(2\pi x)) + \frac{3}{2\pi}\sin(2\pi x) \right ],
\end{equation}
where $K_0$ the zeroth-order modified Bessel function of
the second kind, $a$ the width of the 1D EG ({\it i.e.} the effective
confinement width), and $\epsilon_{\infty}$ is the high frequency
background lattice dielectric constant. 
In the long wavelength limit its asymptotic form becomes
\begin{equation}
V_{\rm c}(q)=\frac{2e^2}{\epsilon_{\infty}}K_0(qa).
\end{equation}
The LO-phonon mediated electron--electron interaction is dependent on
wave vector and frequency, 
\begin{equation}
V_{\rm ph}(q,\omega)=M_{\rm q}^2D_0(\omega).
\end{equation}
$M_{\rm q}$ is the effective 1D Fr\"{o}hlich matrix element given by 
\begin{equation}
M_{\rm q}^2=V_{\rm c}(q)\frac{\omega_{\rm LO}}{2} \left[ 1 -
\frac{\epsilon_{\infty}}{\epsilon_0} \right ],
\end{equation}
where $\omega_{\rm LO}$ is the LO-phonon frequency,
$\epsilon_{0}$ is the static lattice dielectric constant.
The unperturbed retarded bare LO-phonon propagator is given by
\begin{equation}
D_0(\omega)=\frac{2\omega_{\rm LO}}
{\omega^2-\omega_{\rm LO}^2 + i\,0^+\,\mathrm{sgn}(\omega)}.
\end{equation}

The effective electron-electron interaction is obtained in the
RPA by summing (Fig. 1(b)) all the bare bubble diagrams,
\begin{eqnarray}
V_{\rm eff}(q,\omega)&=&\frac{V_{\rm c}(q) + V_{\rm ph}(q,\omega)}{1 -
\left [ V_{\rm c}(q) 
+ V_{\rm ph}(q,\omega) \right ] \Pi_0(q,\omega)} \nonumber \\
&=& \frac{V_{\rm c}(q)}{\epsilon_{\rm t}(q,\omega)},
\end{eqnarray}
where $\Pi_0(q,\omega)$ is the complex irreducible 1D polarizability
given by the bare bubble diagram. The analytic form of
$\Pi_0(q,\omega)$ for complex frequency is given by\cite{li}
\begin{equation}
\Pi_0(q,z)=\frac{m}{\pi q} \ln \left [ \frac{ z^2 - (q^2/2m - q
v_F)^2} {z^2 - (q^2/2m + qv_F)^2} \right ],
\end{equation}
where the principal value of logarithm ({\it i.e.} $ -\pi < 
\mathrm{Im} [\ln(x)]
< \pi$) should be taken. In evaluating $\Pi_0(q,\omega)$ for real
frequency, the limit $z=\omega + i0^+$ is taken.
($v_F$ is the Fermi velocity).
$\epsilon_t(q,\omega)$ is the total dielectric function, which has an
electron and a phonon component,
\begin{equation}
\epsilon_{\rm t}(q,\omega) = 1-V_{\rm c}(q)\Pi_0(q,\omega) +
\frac{1-\epsilon_{\infty}/\epsilon_0}{\epsilon_{\infty}/\epsilon_0 -
\omega^2/\omega_{\rm LO}^2 - i\,\mathrm{sgn}(\omega)\, 0^+ }.
\end{equation}
We include collisional broadening effects arising from
electron-impurity scattering by appropriately modifying\cite{hu} our
dielectric function. Impurity effects in screening are
usually introduced diagrammatically by
including impurity ladder diagrams into the electron-hole
bubble. Since the exact expression for the polarizability within
this ladder vertex diagram approach is complicated, we use the
particle-conserving  
polarizability function, $\Pi_{\gamma}(q,\omega)$,  given by
Mermin\cite{me}. Impurity scattering self-energy 
effects are not included in the single electron Green's
function.  In the presence of an impurity scattering induced level
broadening of $\gamma$, the polarizability function is given by,
\begin{equation}
\Pi_{\gamma}(q,\omega) = \frac {(\omega + i \gamma) \Pi_0(q,\omega +
i\gamma)} {\omega + i\gamma [ \Pi_0(q,\omega+i\gamma)/\Pi_0(q,0)]}.
\end{equation}

\section{ELECTRON SELF-ENERGY}

The electron Green's function $G(k,\omega)$, or equivalently, the
electron self-energy, $\Sigma(k,\omega) = G_0^{-1}(k,\omega) -
G^{-1}(k,\omega)$, where $G_0(k,\omega)$ is the bare noninteracting 
Green's function, provides a complete description of all single
electron properties of an
interacting many-electron system. Knowledge of $G(k,\omega)$ or
$\Sigma(k,\omega)$ allows one to calculate many experimentally
observable one-electron properties of a system, such as the spectral
density function, effective mass, scattering rate, lifetime,
distribution function, band gap renormalization, etc.
The electron self-energy of the coupled system within the
leading-order (Fig. 1(a)) $GW$ 
approximation\cite{hedin2} neglecting vertex corrections at 
$T=0$ is given by
\begin{equation}
\Sigma^t(k,\omega) = i\int^{\infty}_{-\infty} \frac{dq}{2\pi}
\int^{\infty}_{-\infty} \frac{d\omega'}{2\pi} V_{\mathrm{eff}}^t(q,\omega')
G_0^t(k-q,\omega-\omega'),
\end{equation}
where $G_0^t$ is the Green's function for the noninteracting electron gas
(the superscripts $t$ are to denote time-ordering;  
otherwise, the functions in this paper are retarded) 
\begin{equation}
G_0^t(k,\omega) = \frac{1}{\omega-\xi(k)+i0^+{\mathrm{sgn}}(k-k_F)},
\end{equation}
with $\xi(k)=(k^2-k^2_F)/2m$.  The time-ordered quantities are
easily related to the retarded quantities\cite{ma}.
The self-energy can be separated into a frequency-independent 
exchange contribution and a correlation part,
\begin{equation}
\Sigma(k,\omega) = \Sigma_{\rm ex}(k) + \Sigma_{\rm cor}(k,\omega).
\end{equation}
The exchange or Hartree-Fock contribution is given by
\begin{equation}
\Sigma_{\mathrm{ex}}(k) = -\int \frac{dq}{2 \pi} n_F(k+q)V_{\rm c}(q),
\end{equation}
where $n_F(k+q) = \theta(k_F-|k+q|)$ is the Fermi function at $T=0$,
and $\Sigma_{\rm cor}(k,\omega)$ is defined to be the part of
$\Sigma(k,\omega)$ not included in $\Sigma_{\rm ex}(k)$.
In the $GW$ approximation, the $\Sigma_{\rm cor}(k,\omega)$ can be 
written in the line and pole decomposition\cite {ma,qf}
\begin{equation}
\Sigma_{\rm cor}(k,\omega) = \Sigma_{\rm line}(k,\omega) + 
\Sigma_{\rm pole}(k,\omega),
\end{equation}
where
\begin{equation}
\Sigma_{\rm line}(k,\omega) = -\int^{\infty}_{-\infty} \frac{dq}{2\pi}
V_{\rm c}(q)\int_{-\infty}^{\infty} \frac{d\omega'}{2\pi} \frac{1}
{(\xi(k+q)-\omega)-i\omega'} \left[ \frac{1}{\epsilon_t(q,i\omega')}
-1 \right ],
\end{equation}
and
\begin{equation}
\Sigma_{\rm pole}(k,\omega) = \int_{-\infty}^{\infty}\frac{dq}{2\pi}
\left [ \theta\left (\omega-\xi(k+q) \right ) - \theta \left
(-\xi(k+q) \right ) \right]
V_{\rm c}(q) \left [ \frac{1}{\epsilon_t(q,\xi(k+q) - \omega)} - 1 \right ].
\end{equation}
The $\Sigma_{\rm line}(k,\omega)$ is completely real because
$\epsilon_t(q,i\omega')$ is real and even with respect to $\omega'$.
Thus, the imaginary part of the self-energy comes entirely from
Im$[\Sigma_{\rm pole}]$.  
The pole contribution of the self-energy can be written as
\begin{equation}
\Sigma_{\rm pole}(k,\omega) = \left [
\int^{k_0(\omega)}_{k_F}\theta(\omega) -
\int^{k_F}_{0}\theta(-\omega-E_F) - 
\int_{k_0(\omega)}^{k_F}\theta(\omega+E_F)\theta(-\omega) \right ]
F(q,k,\omega) dq,
\label{sigpole}
\end{equation}
where $k_0(\omega) =\sqrt{2m(\omega +E_F)}$ and
$F(q,k,\omega)=\frac{1}{2\pi}[f(q,k,\omega)+f(-q,k,\omega)]$ where
$f(q,k,\omega)$ is given by 
\begin{equation}
f(q,k,\omega) = V_{\rm c}(q-k) \left [
\frac{1}{\epsilon_t(q-k,\xi(q)-\omega)} - 1 \right ].
\end{equation}
Quasiparticle excitation energies $\omega_{\mathrm{q}}$ 
of the system are obtained from the
real part of the self-energy Re$[\Sigma(k,\omega)]$ by solving Dyson's
equation, 
\begin{equation}
\omega_{\mathrm{q}}(k) - \xi(k) - {\rm Re}\Sigma(k,\omega) = 0.
\end{equation}
Once the self energy $\Sigma(k,\omega)$ is known the single particle
spectral function $A(k,\omega)$ is readily calculated. $A(k,\omega)$
contains a important information about the dynamical structure of the
system and is given by
\begin{equation}
A(k,\omega) = \frac{2 |{\rm Im}\Sigma(k,\omega)|} { \left \{ \omega - \xi(k) -
{\rm Re}\Sigma(k,\omega) \right \}^2 + \left \{ {\rm Im} \Sigma(k,\omega)
\right \}^2 }.
\end{equation}
If Im$[\Sigma(k,\omega)]$ goes to zero faster than $|\omega|$ 
as $\omega\rightarrow\omega_{\mathrm{q}}(k)$, the spectral function becomes a 
$\delta$-function peak at the quasiparticle energy $\omega_q$, 
\begin{equation}
A(k,\omega) = Z(k)\; \delta(\omega-\omega_{\mathrm{q}}(k)).
\end{equation}
The strength of the peak is given by the quasi-particle
renormalization factor,
\begin{equation}
Z(k) = \left[1 - \frac{ \partial {\rm Re} \Sigma(k,\omega)}
{\partial \omega}|_{\omega = \omega_{\mathrm{q}(k)}}\right]^{-1}.
\end{equation}
The characteristic of a Fermi system is that it has a finite
discontinuity at the Fermi momentum in the momentum distribution function.
If $Z(k_F) \ne 0$, the electron distribution as a function of momentum
has a finite discontinuity at the Fermi momentum.
Thus, if $A(k_F,\omega)$ shows a $\delta$-function peak 
at $k=k_F$ and $\omega=0$ the system is a Fermi liquid. 
However, if Im[$\Sigma(k_F,\omega)$] goes to zero
slower than $\omega$ as $\omega\rightarrow 0$, 
then the spectral function $A(k_F,\omega=0)$ has
finite broadening at $k_F$ and is not a $\delta$-function, implying
that the system is not a Fermi liquid.

The spectral function $A(k,\omega)$ can be interpreted as a probability
that an electron has a momentum $k$ and energy $\omega$, and satisfies
the sum rule
\begin{equation}
\int^{\infty}_{-\infty} \frac{d\omega}{2\pi} A(k,\omega)=1,
\end{equation}
which is equivalent to the conservation of the number of particles
(electrons) as the interaction is turned on, starting
from a non-interacting picture.
In all our calculations, this sum rule is explicitly checked and is
found to be satisfied to within less than a percent.
The quasiparticle broadening or the damping rate $\Gamma(k)$ is given by the
imaginary part of the self-energy,
\begin{equation}
\Gamma(k) = - {\rm Im} \Sigma(k,\xi(k)),
\end{equation}
where $\xi(k) = k^2/2m - \mu$ is the quasiparticle energy measured
with respect to the chemical potential $\mu$. 
{}From the damping rate we
can calculate the quasiparticle scattering rate $2\Gamma(k)$, the
inelastic life time $\tau(k)=[2\Gamma(k)]^{-1}$, and the inelastic mean free
path $l(k) = v(k) \tau(k)$, where $v(k)$ is the electron velocity.

\section{NUMERICAL RESULTS}

In Fig. 2 we show the real and the imaginary parts of the self-energy
and spectral functions 
for uncoupled (thin lines) and coupled (thick lines) systems
without any impurity scattering. We show the self energies and spectral
functions only at $k=0$ (band edge) and $k=k_F$ (Fermi energy).
We use $E_F$ ($k_F$) as our unit of energy (momentum) throughout our
discussions. 
In this calculation we use the parameters corresponding to 
GaAs: $m=0.07m_{\rm e}$ ($m_{\rm e}$ is the free electron mass),
$\epsilon_0 = 12.9$, $\epsilon_{\infty}=10.9$, and $\omega_{\rm LO} =
36.8$\,meV.  The well width of $a=100 \AA$ and the electron gas density 
of $n=0.56 \times 10^6 {\rm cm}^{-1}$, which corresponds to a
Fermi energy $E_F\approx 4.4$meV and a dimensionless density parameter
$r_{\rm s}=4me^2/\pi k_F \epsilon_0 = 1.4$ with $k_F = \pi n/2$, are
used for this calculation. 
The intersections of Re$[\Sigma(k,\omega)]$ and the  straight line
$\omega - \xi(k)$ indicate the solutions to Dyson's equation. 
For $k=0$ and the uncoupled system we find three solutions of the
Dyson's equation (Eq. 18). Two solutions (the first and the third one
from $\omega=0$) show the $\delta$-function peaks
in the spectral function, but the second solution has a broad
incoherent structure indicating finite damping in the spectral
function. The first 
$\delta$-function peak in $A(k,\omega)$ (with $Z=2\pi \times 0.33$),
which corresponds to the usual quasi-particle 
(i.e., a bare particle surrounded by a cloud of virtual plasmons and
particle-hole excitations), is slightly shifted from the noninteracting
single particle energy $\omega=\xi(k)$. The second $\delta$-function
peak (with $Z=2\pi \times 0.31$) has been called a plasmaron which is
interpreted as a hole coupled to a cloud of real plasmons.
For the Fermi wave vector ($k=k_F$) there is a only one solution to
Dyson's equation at $\omega=0$ and a strong peak in $A(k,\omega)$.
However, this peak is not the $\delta$-function peak, since
Im$[\Sigma(k_F,\omega)]\rightarrow \omega \sqrt{\ln \omega}$ as
$\omega \rightarrow 0$\cite{hu}. Thus the uncoupled 1D EG without any
impurity scattering is not a Fermi liquid even within the
leading-order GW approximation\cite{hu}, and does not have
well-defined quasiparticles. This was already reported in
ref. \onlinecite{hu}. 

Since Im$[\epsilon_c(q,\omega)]^{-1}\neq0$ (the subscript $c$ stands
for the Coulomb part; i.e., the uncoupled system) within the electron-hole
continuum, the contribution to 
the imaginary part of the self-energy, Im$[\Sigma(k,\omega)]={\rm
Im}[\Sigma_{\rm pole}(k,\omega)$], comes from
both the electron-hole continuum and the collective plasmon mode. 
Im$[\epsilon_c(q,\omega)]^{-1}$ can be written as 
\begin{eqnarray}
{\rm Im}\frac{1}{\epsilon_c(q,\omega)}& = &\frac{-\epsilon_2}{\epsilon_1^2 +
\epsilon_2^2} \left [ \theta(|\omega_1|-\omega)\theta(\omega -
|\omega_2|) \theta(\omega) + \theta(-|\omega_2|-\omega)\theta(\omega +
|\omega_1|) \theta(-\omega) \right ] \\ \nonumber
                            &   & - \pi \left [ \frac{\partial
\epsilon_1}{\partial \omega}|_{\omega=\omega_{\rm pl}}  \right ] \delta(\omega
- \omega_{\rm pl}),
\label{imeps}
\end{eqnarray}
where $\epsilon_1={\rm Re}[\epsilon(q,\omega)]$ and $\epsilon_2={\rm
Im}[\epsilon(q,\omega)]$, and the plasmon dispersion, $\omega_{\rm pl}$,
is given by\cite{li}
\begin{equation}
\omega_{\rm pl}(q)=\left [ \frac{g(q)\omega_1^2 - \omega_2^2} {g(q) -
1} \right ]^{1/2},
\end{equation}
where $g(q)=\exp[q\pi/mV_{\rm c}]$, $\omega_1 = 2q + q^2$, and
$\omega_2 = 2q -q^2$.
The first (second) term in Eq. (25) gives the electron-hole
(plasmon) contribution. 
{}From Eq. (\ref{sigpole}), we see that the non-zero electron-hole
contribution shows up in 
Im$[\Sigma(k,\omega)]$ only if $\omega > 1-k^2$
for $\omega > 0$ and if $-(k+1)(k+3) < \omega < -|1-k^2|$ for
$\omega<0$. The plasmon part (the second term in Eq. (25) 
contributes to Im$[\Sigma(k,\omega)]$ if $\omega > \omega_{\rm pl}(1 \pm k)$
for  $\omega > 0$ and if $-\omega_{\rm pl}(1\pm k) < \omega <
-|1-k^2|$ for $\omega<0$. 
Thus, for $k=0$, Im$[\Sigma(k,\omega)]\neq
0$ in the range $-\omega_{\rm pl}(1) < \omega < -1$ and $\omega >1$ 
(the electron-hole contribution, $\omega>1$ and $-3 < \omega <-1$, and
the plasmon contribution, $-\omega_{\rm pl}(1) < \omega < -1$ and $
\omega > \omega_{\rm pl}(1)$).
Since the plasmon contributions turn on at $\omega=\pm\omega_{\rm pl}(1)$ 
the finite discontinuities in Im$[\Sigma(k,\omega)]$ take place at
those points 
with the magnitude $\pi/\frac{\partial \epsilon_1}{\partial
\omega}|_{\omega = \pm \omega_{\rm pl}}$. A finite discontinuity in
Im$[\Sigma]$ gives rise to a logarithmic singularity in Re$[\Sigma]$,
which can be verified using the Kramers-Kronig relation. 
For $k=k_F$, Im$[\Sigma] \neq 0$ in the range $\omega > 0$ and
$-\omega_{\rm pl}(2) < \omega <0$ ( the electron-hole contribution,
$\omega > 0$ and $-6<\omega<0$, and the plasmon contribution, $\omega
> \omega_{\rm pl}(2)$ and $-\omega_{\rm pl}(2) < \omega <0$). Turning
on the plasmon contribution at $\omega=\pm \omega_{\rm pl}(2)$ induces
the finite discontinuity in Im$[\Sigma]$ and a logarithmic singularity
in Re$[\Sigma]$.

For the coupled 1D EG and $k=0$ we find three undamped solutions and 
two damped solutions in Dyson's equation. 
The first $\delta$-function peak in
$A(k,\omega)$ near $\omega=0$, the quasi-particle peak, occurs 
at almost the same energy as the uncoupled case, but with slightly
higher strength ($Z=2\pi \times 0.36$).
The second $\delta$-function peak, the plasmaron peak, is shifted
toward the Fermi energy and has much weaker strength ($Z=2\pi
\times0.22 $) than in the uncoupled 1D EG, since the coupled
plasmon-like mode has a 
smaller energy and weaker strength than in the uncoupled case\cite{hwang}.
Unlike the uncoupled system we now find a 
third $\delta$-function peak with strength $Z=2\pi \times 0.05$ near
the frequency $\omega=-E_F-\omega_{\rm LO}$. This
phonon satellite can be interpreted as being a ``phononron'' similar
to the plasmaron, that is, a 
quasi-hole coupled to real LO-phonons. For $k=k_F$ we 
find only two peaks : the quasi-particle peak at $\omega=0$ and the
phonon satellite 
(the $\delta$-function peak with strength $Z=2\pi \times0.03 $). 
As $\omega \rightarrow 0$ the main contribution to Im$[\Sigma]$ comes
from a plasmon, since the phonon part has a gap at $q=0$ and the
integration is performed near $q=0$. Thus, the behavior of
Im$[\Sigma]$ for the coupled system is the same as that for
the uncoupled system and the quasi-particle like peak near $\omega=0$
is not a strict $\delta$-function peak, which means that the coupled
1D EG with electron-phonon interaction is not a Fermi liquid either,
and has no true long-lived quasi-particles.

For the coupled system there is also a direct phonon contribution to the
imaginary part of the self-energy along with the electron-phonon and plasmon
contribution discussed above.  Im$[\epsilon_{\rm t}]^{-1}$ can be written as 
\begin{eqnarray}
{\rm Im}\frac{1}{\epsilon_t(q,\omega)}& = &\frac{-\epsilon_2}{\epsilon_1^2 +
\epsilon_2^2} \left [ \theta(|\omega_1|-\omega)\theta(\omega -
|\omega_2|) \theta(\omega) + \theta(-|\omega_2|-\omega)\theta(\omega +
|\omega_1|) \theta(-\omega) \right ] \\ \nonumber
                            &   & - {\pi}\left [ \frac{\partial
\epsilon_1}{\partial \omega}|_{\omega=\omega_{\rm pl}^{(-)}} \right
]^{-1} \delta(\omega
- \omega_{\rm pl}^{(-)}) - \pi \left [ \frac{\partial
\epsilon_1}{\partial \omega}|_{\omega=\omega_{\rm pl}^{(+)}}\right
]^{-1} \delta(\omega 
- \omega_{\rm pl}^{(+)}),
\end{eqnarray}
where $\omega_{\rm pl}^{(-)}$ and $\omega_{\rm pl}^{(+)}$ are the
plasmon-like and the phonon-like collective modes in the 
coupled system, respectively\cite{hwang}. Their behaviors in the long
wave length limit ($q \rightarrow 0$) are given by
\begin{equation}
\omega_{\rm pl}^{(+)}(q)=\omega_{LO} \left [1+\frac{\omega_{LO}^2 -
\omega_{TO}^2} {\omega_{LO}^4}2r_s|\ln(qa)|q^2 + O(q^4) \right ],
\end{equation}
and
\begin{equation}
\omega_{\rm pl}^{(-)}(q) = \frac{\omega_{TO}}{\omega_{LO}}2q\sqrt{r_s
|\ln(qa)|} + O(q^3).
\end{equation}
The contribution to Im$[\Sigma]$ from the electron-hole continuum is the
same as that for the uncoupled 
system, but the contribution of the plasmon-like mode, $\omega_{\rm
pl}^{(-)}$, is the same only as long as  $\omega < \omega_{\rm TO}$, since the
plasmon-like mode ceases for $\omega > \omega_{\rm TO}$ [See Ref.
\onlinecite {hwang}]. The
discontinuities in Im$[\Sigma]$ induced by this mode move toward the
Fermi energy ($\omega=0$) and the strength of the step becomes
smaller, since $|\omega_{\rm pl}| > |\omega_{\rm pl}^{(-)}|$. The 
phonon-like mode contributes to Im$[\Sigma]$ if the following
conditions are satisfied: if $\omega > 0$, $q^2 - 1 - \omega +
\omega_{\rm pl}^{(+)} = 0$ with $1 < q < \sqrt{\omega + 1}$, and if
$\omega < 0$, $q^2 -1 -\omega -\omega_{\rm pl}^{(+)} = 0$ with $0 < q
< 1$. From these conditions we find the range in which the phonon-like
mode contributes to the Im$[\Sigma]$ to be $\omega > \omega_{\rm
pl}^{(+)}(1\pm k)$ and $-1-\omega_{\rm pl}^{(+)}(k) < \omega <
-\omega_{\rm pl}^{(+)}(1- k)$. As for the plasmon, turning on 
the phonon-like mode produces a discontinuity in Im$[\Sigma]$ and a
singularity in Re$[\Sigma]$. For $k=k_F$, since $\omega_{\rm
pl}^{(+)}(1\pm k) \geq \omega_{\rm TO}$, the abrupt steps in Im$[\Sigma]$
for $\omega > 0$ are due to the phonon-like mode. 

Fig. 3 shows the results for the coupled 1D EG with impurity scattering
effects ($\gamma = E_F$). The inclusion of
impurity scattering introduces collisional damping in the collective
modes. Since the 1D
plasmon is strongly affected by impurity scattering, (i.e., they become 
overdamped at small $q$), the damping of the plasmaron 
(the second peak in $A$ from $\omega=0$) peak is stronger than that 
for other excitations.  However, the phonon part of the dynamical 
dielectric function is not affected by the impurity scattering. Thus,
the phonon satellite (the third peak from $\omega=0$ in
$A$) still shows a sharp peak ( almost undamped ), and the abrupt steps
in Im$[\Sigma]$ still exist at $\omega=\pm \omega_{\rm LO}$ and
$\omega=-E_F - 
\omega_{\rm LO}$. The discontinuities in Re$[\Sigma]$ and the
singularities in Re$[\Sigma]$ induced by the plasmon-like mode are
both suppressed
by the impurity scattering. For $k=k_F$, the contribution to
Im$[\Sigma(k=k_F,\omega)]$ at $\omega=0$ mainly comes from the single
electron-hole pair excitations as in the uncoupled 1D EG, since the
long wavelength plasmon is now
overdamped and the phonon has a gap at a small $q$. 
As $\omega \rightarrow 0$ the contribution to Im$[\Sigma]$ from the 
single-particle excitations is given by $\omega^2|\ln \omega|^3$,  and
$A(k_F,\omega) \sim |\ln(|\omega|)|^{3} + 2\pi Z_{F}
\delta(\omega)$\cite{hu}.
Our direct numerical calculation shows that the weights of the $\delta$
function are $2 \pi \times 0.31$ for the uncoupled system and $2 \pi
\times 0.33$ for the coupled system.
Therefore, the coupled 1D EG with impurity scattering is a Fermi
liquid similar to the behavior found in ref. \onlinecite{hu} for the
uncoupled system in the presence of impurity scattering.

In Fig. 4 we show the strength $Z(k)$ of the undamped excitations for
uncoupled (thin lines) and coupled (thick lines) system without
impurity scattering  
as a function of wave vector. As the
wave vector increases the strength of the $\delta$-function peaks for
the plasmarons (dashed lines) decrease quickly and vanish at
$k=0.75k_F$ for the 
uncoupled system and at $k=0.3k_F$ for the coupled system, but the
strengths of the regular quasiparticle peaks (solid lines) and the
phononron peak (dot-dashed line)
decrease slowly. On the other hand, as 
the wave vector approaches the Fermi momentum the strength of the
quasiparticle peak decreases and vanishes at the Fermi momentum.
Analytically we can show that this behavior is logarithmic as in the  
uncoupled system. The most dominant term in the derivative of the real
part of the self-energy comes from $\Sigma_{\rm pole}$ as $\omega
\rightarrow 0$. From Eq. (16) we have as $\omega \rightarrow 0$
\begin{eqnarray}
\frac {\partial \Sigma_{\rm pole}} {\partial \omega}|_{\rm k=k_{F}} 
& = & \frac {1}{2\pi}\int_{-\infty}^{\infty} dq \delta(\omega -
\xi_{\rm k+q}) v_{\rm c}(q) \left [ \frac{1}{\epsilon_t(q,\xi_{\rm k+q}
- \omega)} - 1 \right ] \nonumber \\
& = & \frac {1}{2 \pi} \left [ v_{\rm c}(\frac {\omega}{2}) \left \{
\frac{1}{\epsilon_t(\frac{\omega}{2},0)} -1 \right \} + v_{\rm c}(2)
\left \{ \frac{1}{\epsilon_t(2,0)} -1 \right \} \right ].
\end{eqnarray}
Thus, as $\omega \rightarrow 0$ the first term shows a logarithmic 
singularity,
\begin{equation}
\frac {\partial \Sigma_{\rm pole}} {\partial \omega}|_{\rm k=k_{F}} 
\sim - \left ( \ln|\omega| + 1 \right ).
\end{equation}
This demonstrates that even within our leading order approximation the
coupled system without impurity scattering is not 
a Fermi liquid, since the renormalization constant at $k=k_F$
vanishes.
(i.e., the momentum distribution is continuous at the Fermi momentum).

Fig. 5(a) shows the quasiparticle damping rate $\Gamma(k)$ for 
both the coupled system (thick lines) and the
uncoupled system (thin lines) with the parameters $a=100 \AA$
and $n=0.56 \times 10^6 cm^{-1}$. (Inset in
Fig. 5(a) shows the plasmon-phonon mode coupling.) 
The corresponding inelastic mean free paths $l_K$ are shown in Fig
5(b). For the uncoupled 1D EG the quasiparticle scatters by plasmon
emission, which corresponds to the sharp threshold in the thin
solid line shown in Fig. 5(a). There is no 1D single particle
electron-electron scattering below a threshold critical wave vector
because the conservation of energy and momentum restricts
electron-electron scattering to an exchange of particles.
In 2D and 3D systems there is allowed
scattering below the threshold critical wave vector due to the
excitation of electron-hole pairs.  For the coupled 1D EG the
quasiparticle decays via the emission of coupled plasmon-phonon mode\cite{kim}
($\omega_-$ and $\omega_+$ in the inset of Fig. 5(a)), which
corresponds to the two peaks in the thick solid line of Fig. 5(a). The
first step corresponds to the $\omega_-$-emission threshold and is
located below the $k_c$ of the uncoupled 1D EG because $\omega_- <
\omega_0$. The second step corresponds to the $\omega_+$-emission
threshold and occurs at a wave vector larger than $k=\sqrt{\omega_{LO}
+ 1}$ because $\omega_+ > \omega_{LO}$. 
When only Fr\"{o}hlich interaction in the 1D quantum wire is 
considered\cite{rid} the LO-phonon emission threshold has been predicted at
the fixed critical wave vector $k=\sqrt{\omega_{LO}+1}$. But in our
coupled model the locations of $\omega_{\pm}$-emission threshold depend on
the density of the electron as in the plasmon emission threshold.
The thin lines in Fig. 5(a) show
the damping rate with the impurity scattering effects
($\gamma=E_F$). Since the plasmon line is broadened by impurity scattering
there are inelastic scattering events below the critical wave vectors.
The breaking of translational invariance by impurities, furthermore, relaxes
momentum conservation and permits inelastic scattering via
single-particle excitation (except at the Fermi surface where the
quasiparticle is always undamped). Thus, the nature of the sharp
threshold in our doped system is totally different from that discussed
in ref. \onlinecite{rid} which considers undoped quantum wires.
The sharp divergent thresholds obtained in Ref. \onlinecite{rid} 
result from the divergence in the density of states at the bottom 
of the band of one-dimensional systems, and therefore in doped samples,
when the bottom of the band is no longer accessible, these sharp
thresholds do not occur.  In our theory, the divergences are
due to the divergence in the joint density of states at the
coupled plasmon-phonon emission threshold, and they occur 
at any doping.

\section{SUMMARY and CONCLUSION}

In this paper we have calculated the quasiparticle self-energy, the
spectral function, and the damping rate within the leading order
dynamical screening approximation by treating electron-electron and
electron-phonon interactions on an equal footing both with and without
impurity scattering effects. Our many-body theory includes the
important physical effects of
dynamical screening, phonon self-energy correction, plasmon-phonon
mode coupling, Fermi statistics, and Landau damping. Even though the
problem is treated within the $GW$ framework of the leading order effective 
interaction approximation (i.e. RPA screening and neglect of vertex
corrections), our results should be quite valid in GaAs-based 1D EG.
This is because GaAs has a very weak Fr\"{o}hlich coupling, 
which justifies the neglect of the electron-phonon vertex corrections, 
and its low effective mass and the large dielectric constants 
gives a small effective $r_s$ parameter, (and in this limit generally 
RPA is valid), which makes direct electron-electron interaction also
weak in the perturbative sense.

The quasiparticle properties at $\omega\approx \xi_k$ are affected
only by low energy processes and hence are not changed much by the 
inclusion of the optic-phonon coupling because the optic-phonon modes 
have a gap to excitation.  The optic-phonon coupling does, however,
shift some of the spectral weight to an extra undamped excitation at 
$\omega \approx -(E_F + \omega_{LO})$,   
which is the phonon analogy (``phononron'') to the plasmaron excitation,
and it shifts the plasmaron energy up slightly, due to level repulsion
between the plasmaron and the phonon excitation.  
As in the uncoupled system, inclusion of impurity scattering tends 
to smear out the above features.

The properties at the Fermi surface, such as the impurity-driven 
restoration of the discontinuity in the distribution
function, are virtually the same for the coupled and uncoupled systems,
again because of the gap of the optic-phonon mode.  
At higher energies, the coupled system has an additional phonon-like 
mode $\omega_+$ which can scatter electrons.  This leads to another
divergence in the scattering rate $\Gamma(k)$ (i.e., 
in addition to the one from plasmon emission).  
Note that this purely many-body
divergence occurs only when one couples the electron--electron
and electron--phonon interactions together; no such divergence is
obtained for calculations in doped systems using only the Fr\"{o}hlich
interaction.

\section*{ACKNOWLEDGMENTS}
This work is supported by the U.S.- ARO and U.S.- ONR.

\begin{figure}
\caption{(a) Electron self-energy in leading order in the effective dynamical
interaction. (b) Effective dynamical interaction (thick wiggly line)
$V_{\rm eff}$ calculated in the RPA. Thin wiggly lines (dashed
lines) represent the Coulomb electron-electron interaction $V_{\rm c}$ (the
LO-phonon-mediated electron-electron interaction $V_{\rm ph}$), and the
bubble the irreducible polalizability $\Pi_0$.
(c) Higher-order self-energy diagrams neglected in our calculation.}
\end{figure}

\begin{figure}
\caption{ Self energy $\Sigma(k,\omega)$ ( (a) for $k=0$ and (b) for
$k=k_F$ ) and 
spectral function $A(k,\omega)$ ( (c) for $k=0$ and (d) for $k=k_F$ )
as functions of the 
frequency $\omega$ without any
impurity scattering ($\gamma=0$). 
The straight lines are given by $\omega - \xi_{\rm k} - \mu$, and
their interactions with Re[$\Sigma$] indicate the solutions to Dyson's
equation and correspond to a quasiparticle peak. For clarity we plot
$|{\rm Im}\Sigma|$. Thin (thick) lines
correspond to the uncoupled (coupled) 1D EG. The vertical arrows in
(c) and (d) denote $\delta$ functions in the spectral function.
For $k=0$, we find  at $\omega = $ -4.85
and -0.96 with weights $2\pi \times 0.31$ and $2 \pi \times 0.33$
respectively for the uncoupled system, and at $\omega =
$ -9.65, -4.28, and -0.96 with weights $2\pi \times 0.05$, $2 \pi
\times 0.22$, and $2 \pi \times 0.36$ respectively for the coupled system.
For $k=k_F$, we find a peak at $\omega=-9.81$ with weight $2\pi \times
0.03$ only for the coupled system.
The parameters corresponding to 
GaAs are used: $m=0.07m_{\rm e}$ ($m_{\rm e}$ is the free electron mass),
$\epsilon_0 = 12.9$, $\epsilon_{\infty}=10.9$, and $\omega_{\rm LO} =
36.8$ meV.  The well width of $a=100 \AA$ and the
electron gas density of $n=0.56 \times 10^6 {\rm cm}^{-1}$, which
corresponds to a 
Fermi energy $E_F\approx 4.4$meV and a dimensionless density parameter
$r_{\rm s}=4me^2/\pi k_F \epsilon_0 = 1.4$ with $k_F = \pi n/2$, are
used for this calculation. 
}
\end{figure}

\begin{figure}
\caption{ Self energy $\Sigma(k,\omega)$ ( (a) for $k=0$ and (b) for
$k=k_F$ ) and 
spectral function $A(k,\omega)$ ( (c) for $k=0$ and (d) for $k=k_F$ )
as functions of the 
frequency $\omega$ with
impurity effects ($\gamma=E_{\rm F}$). Thin (thick) lines
correspond to the uncoupled (coupled) 1D EG. For $k=k_{\rm F}$,
vertical arrows in the spectral  function at $\omega=0$ 
denote $\delta$
functions with weight $2\pi \times 0.31$ for uncoupled system
and $2\pi \times 0.33$ for uncoupled system. }
\end{figure}

\begin{figure}
\caption{Strength of the undamped excitations (the
quasiparticle renormalization factor $Z(k)$) 
for uncoupled (thin lines) and  
coupled (thick lines) system without impurity scattering as a function
of wave vector.  
Solid lines (dashed lines) denote quasiparticle
peaks (plasmarons) and dot-dashed line a phononron. Inset
shows the logarithmic approach of the 
renormalization factor to zero as $k \rightarrow k_{\rm F}$.}
\end{figure}

\begin{figure}
\caption{(a) Damping rate $\Gamma(k)$ and (b) the
corresponding mean free path $l(k) = v(k)/2\Gamma(k)$ as a function of
$k$. Solid (dashed) lines indicate the coupled (uncoupled) system and
thick (thin) lines for $\gamma = 0$ 
($\gamma = E_F$).  The inset in (a) shows the plasmon-phonon mode coupling.}
\end{figure}


\begin{thebibliography}{99}
\bibitem{typo} Note that Eq.\ (2.5) in Ref. \protect\onlinecite{hu}
contains typographical errors which are corrected here.
\end{thebibliography}
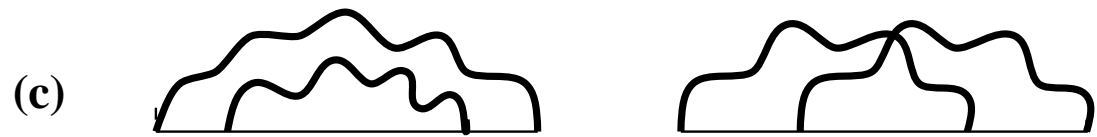
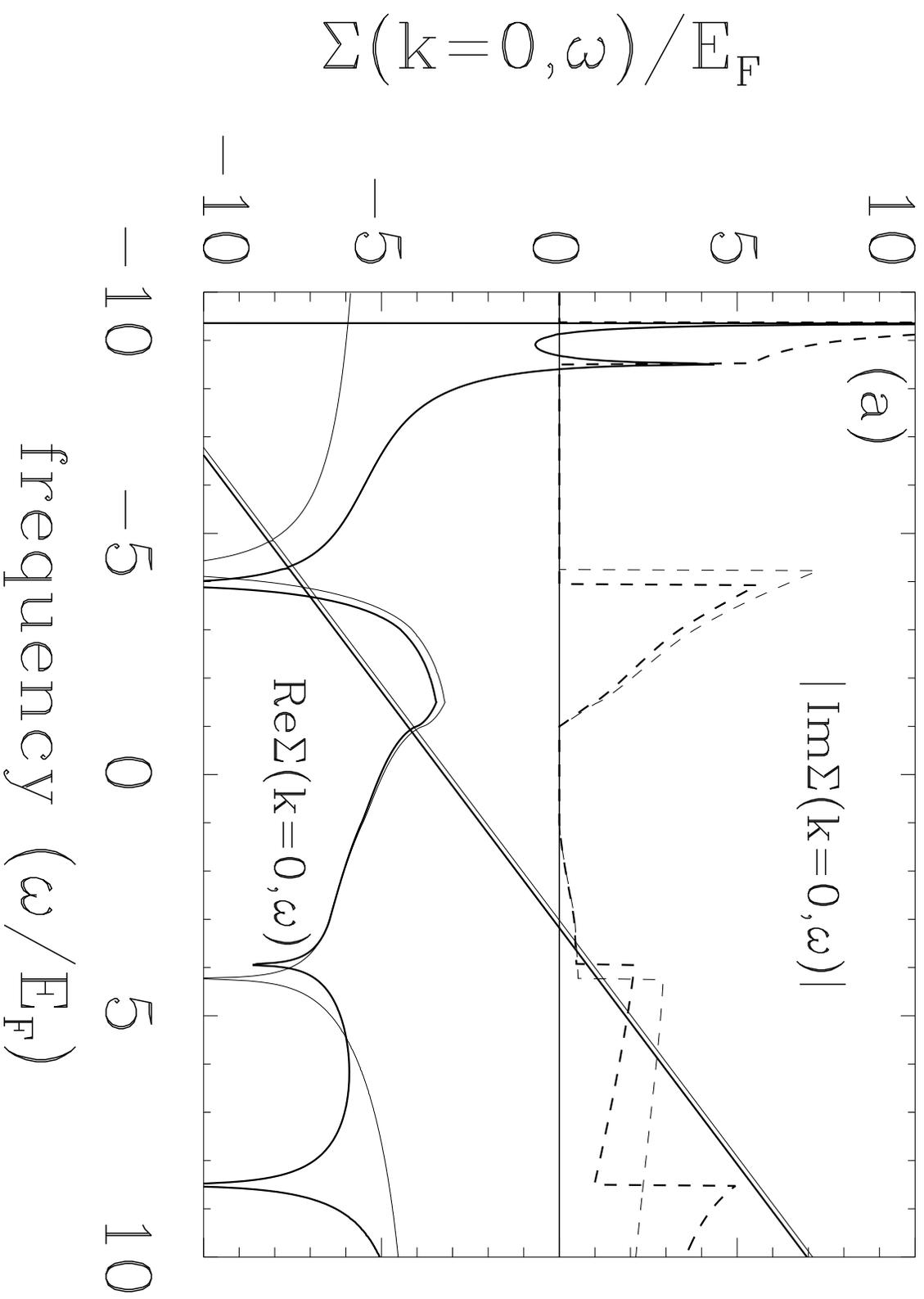
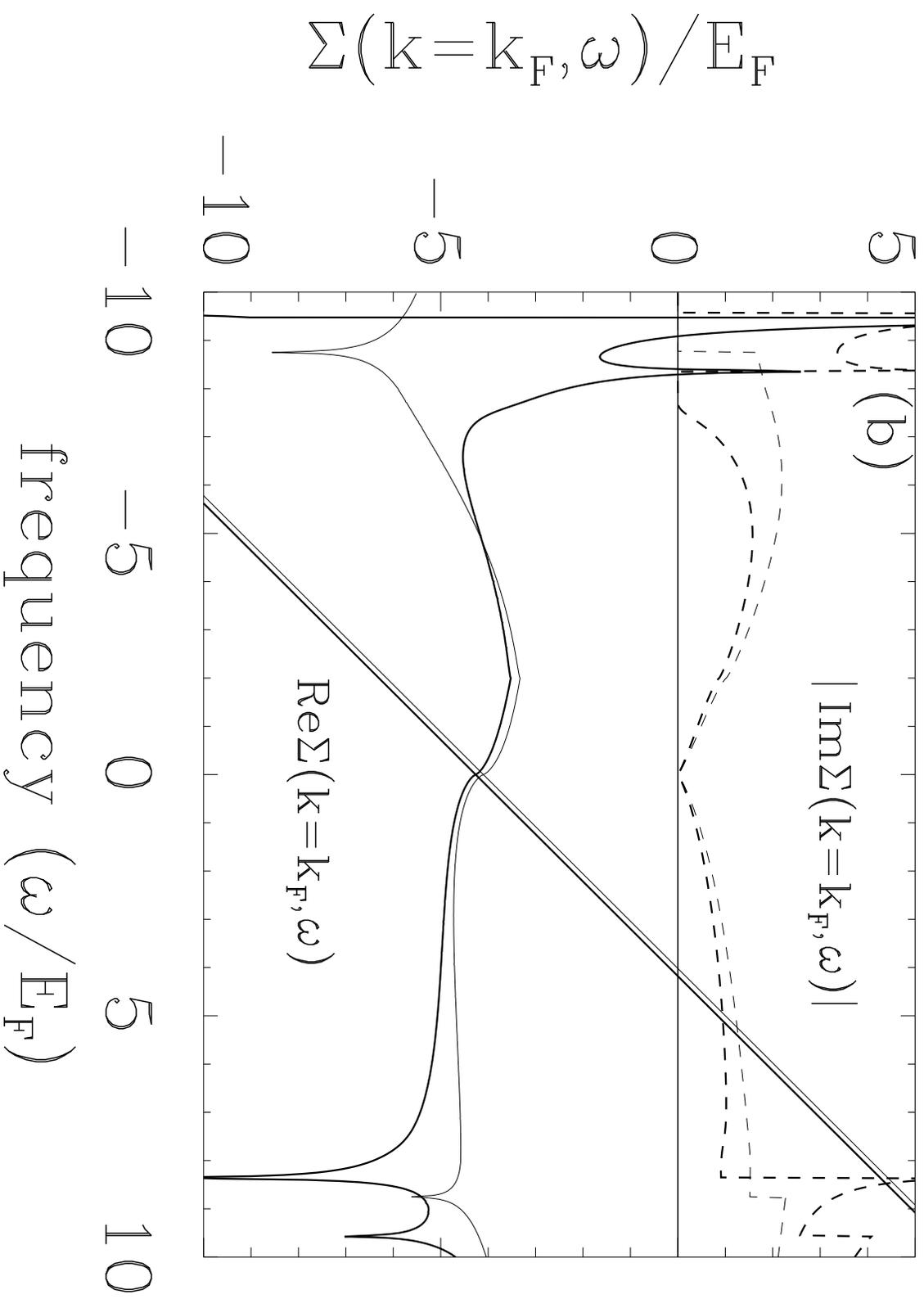
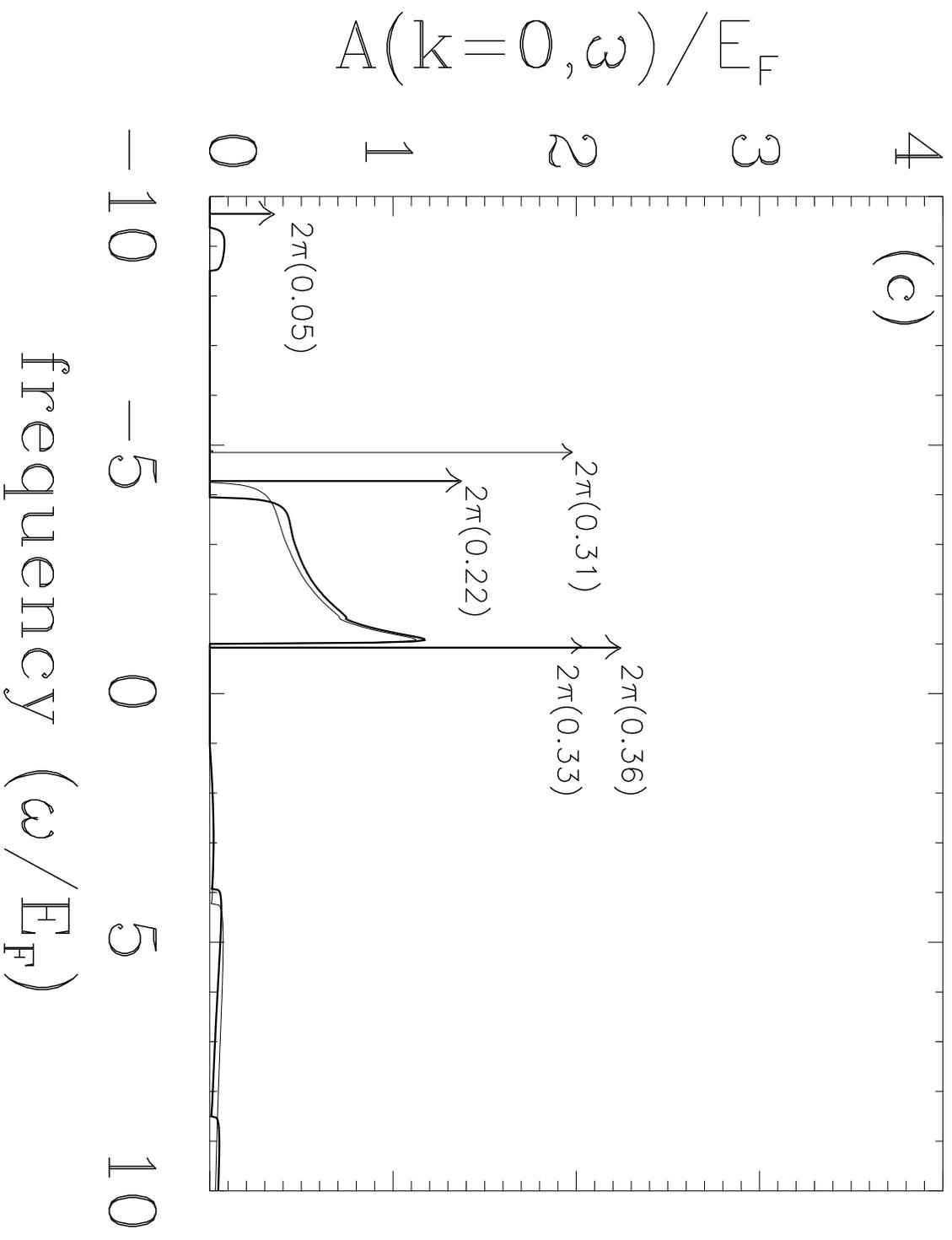
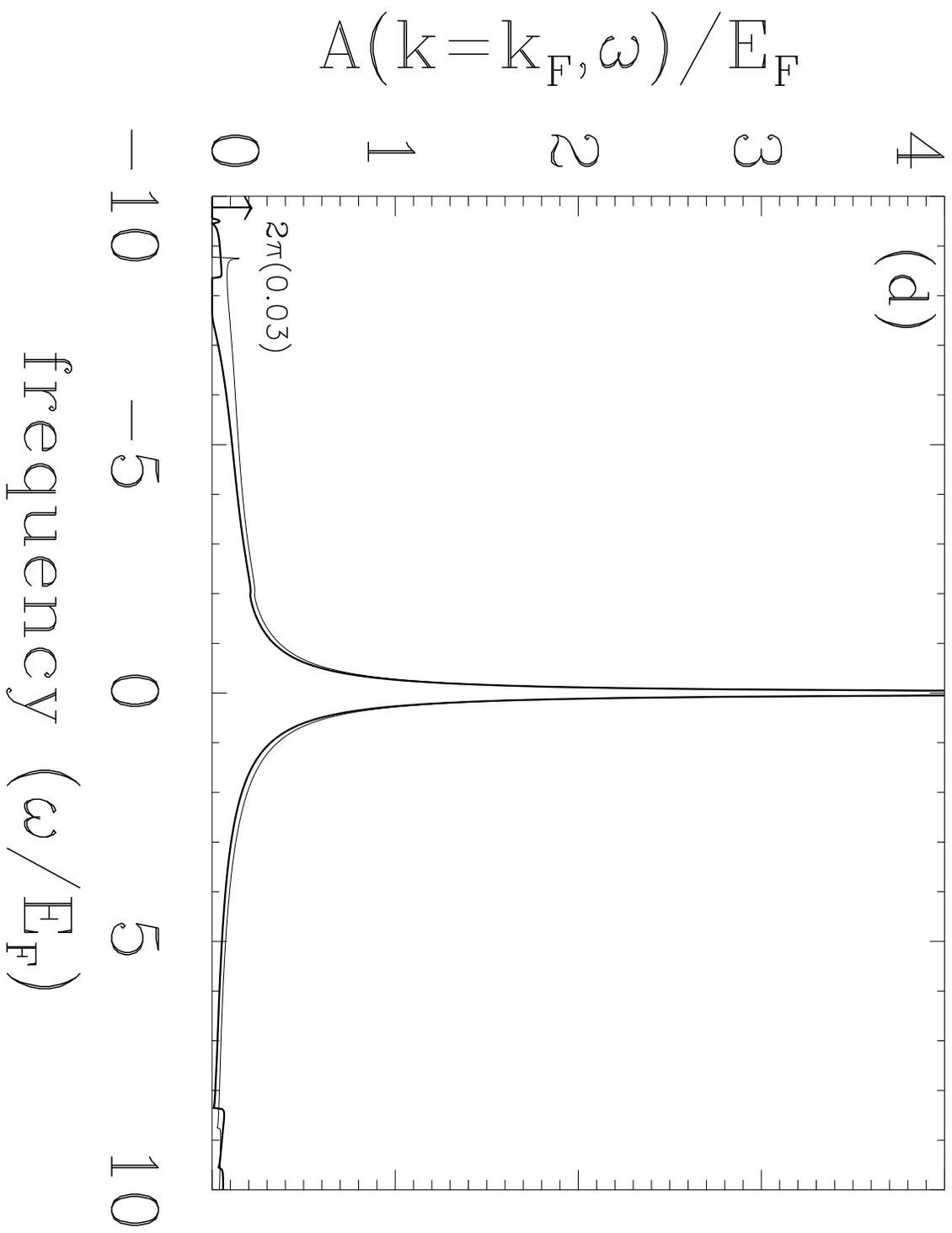
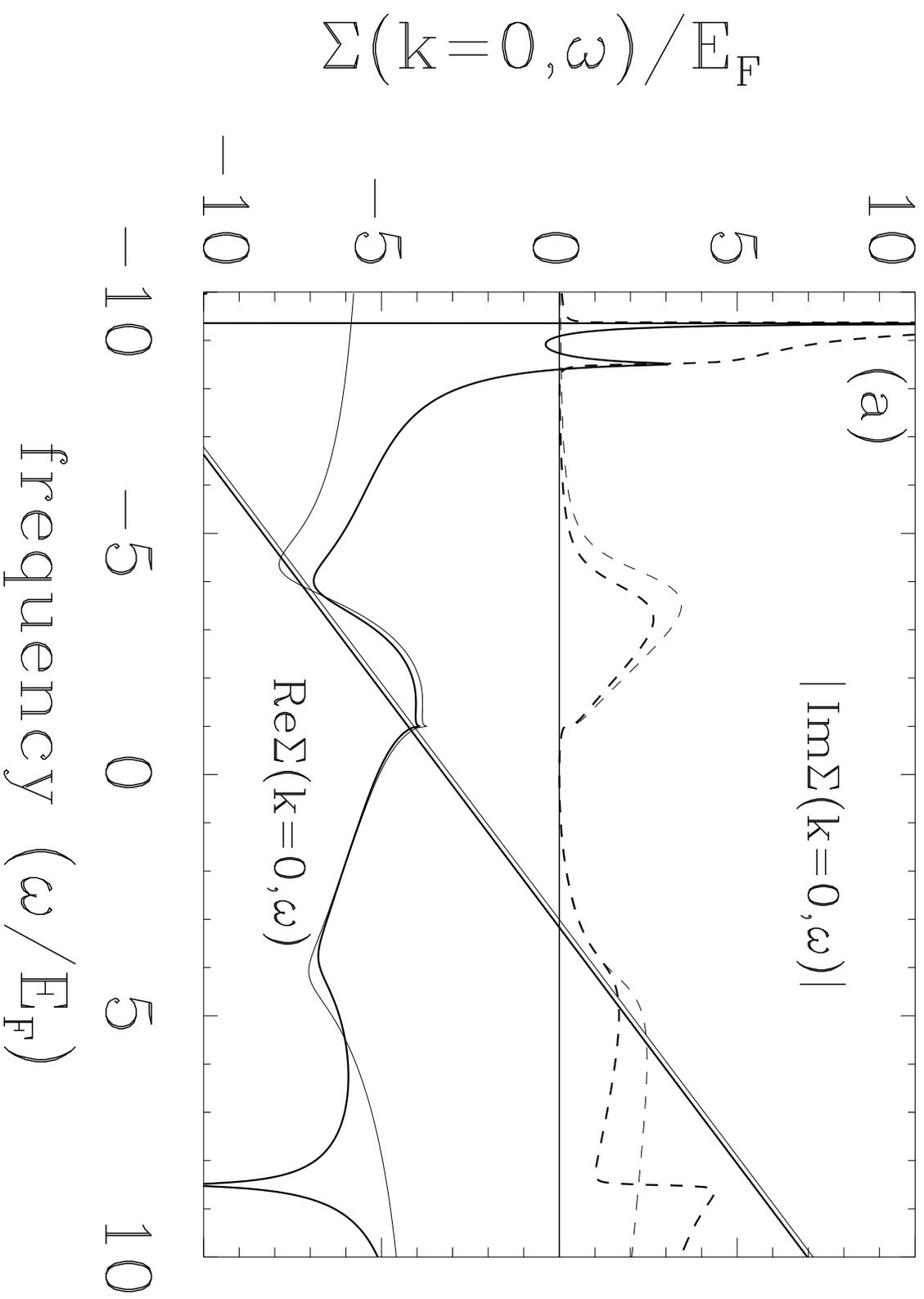
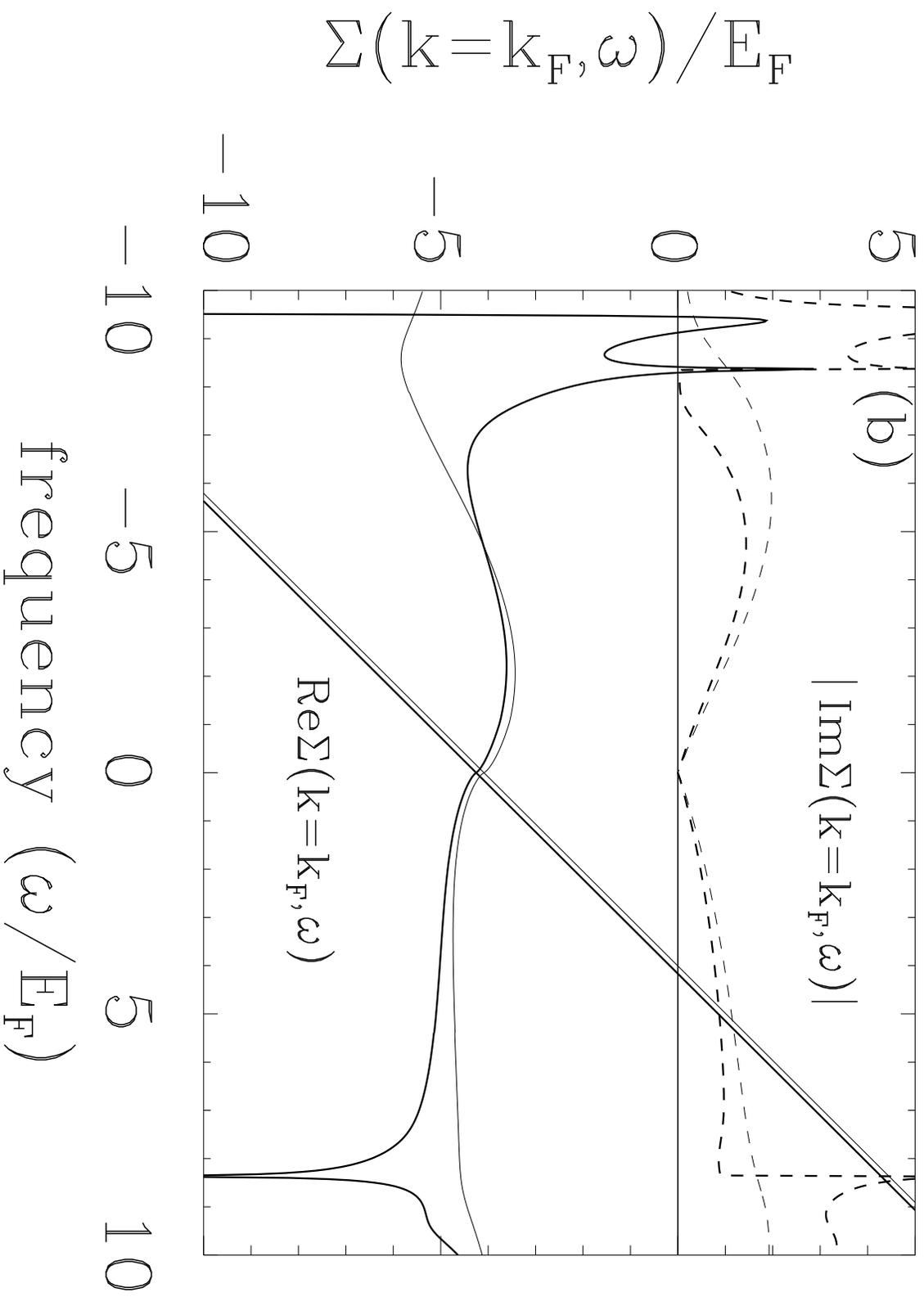
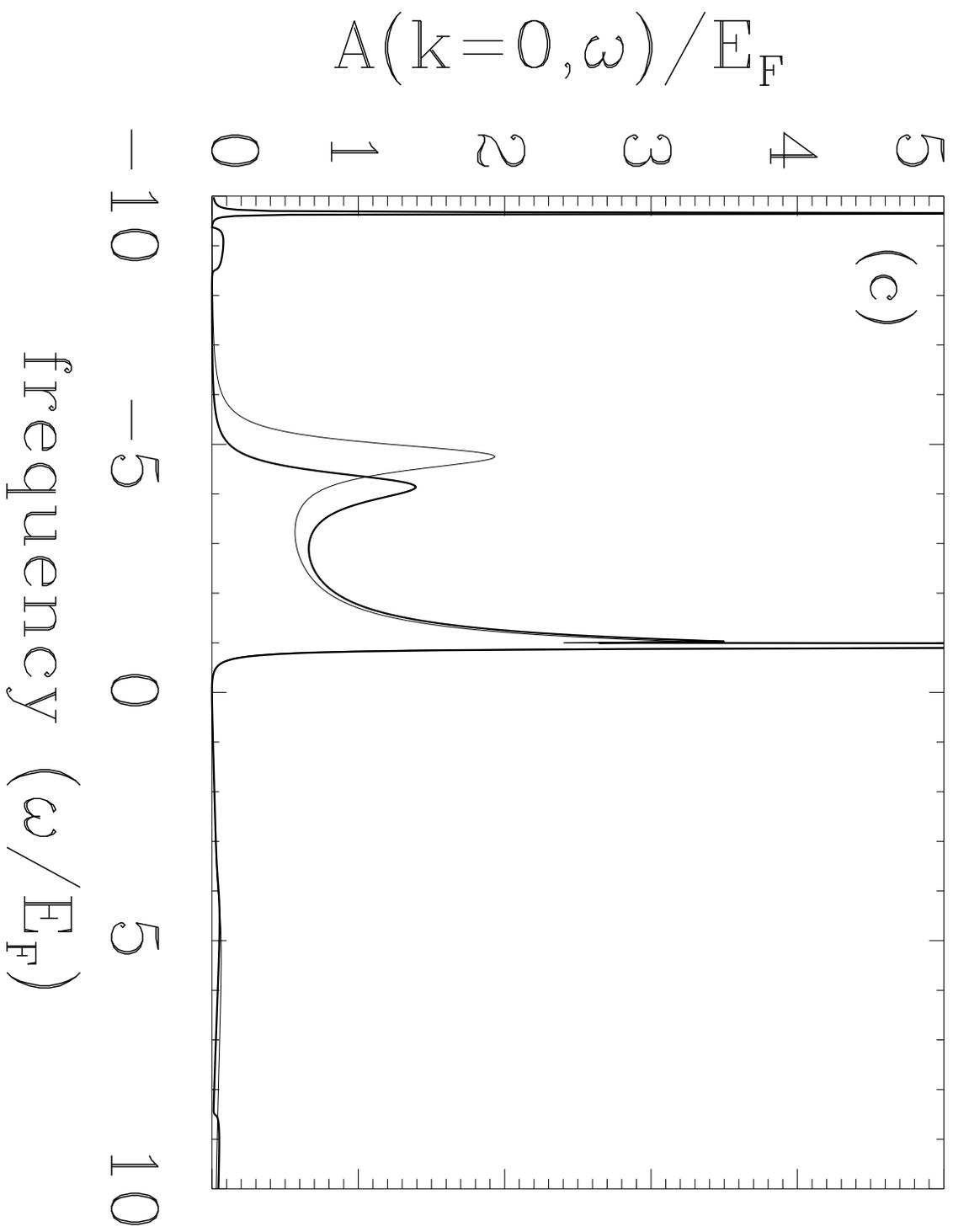
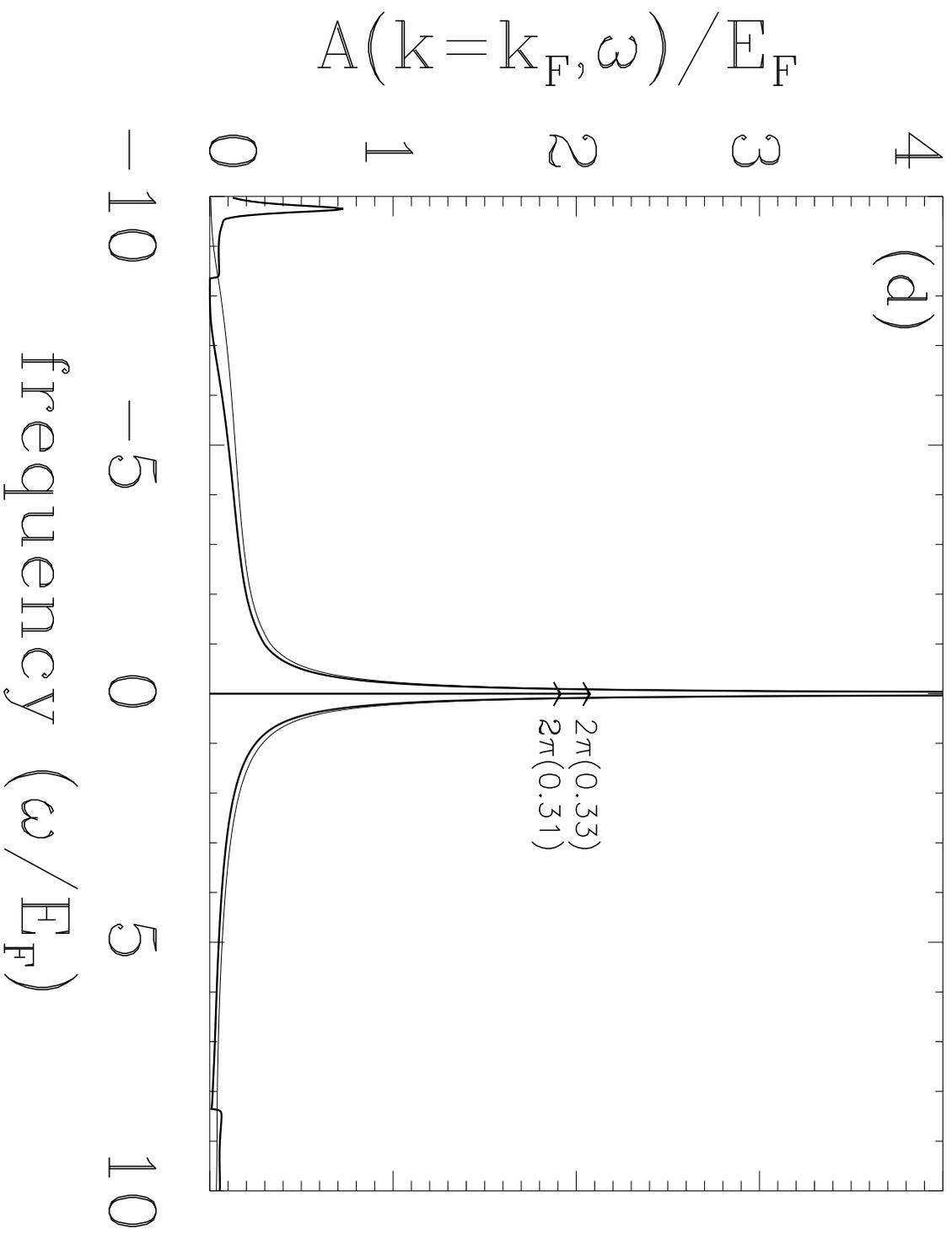
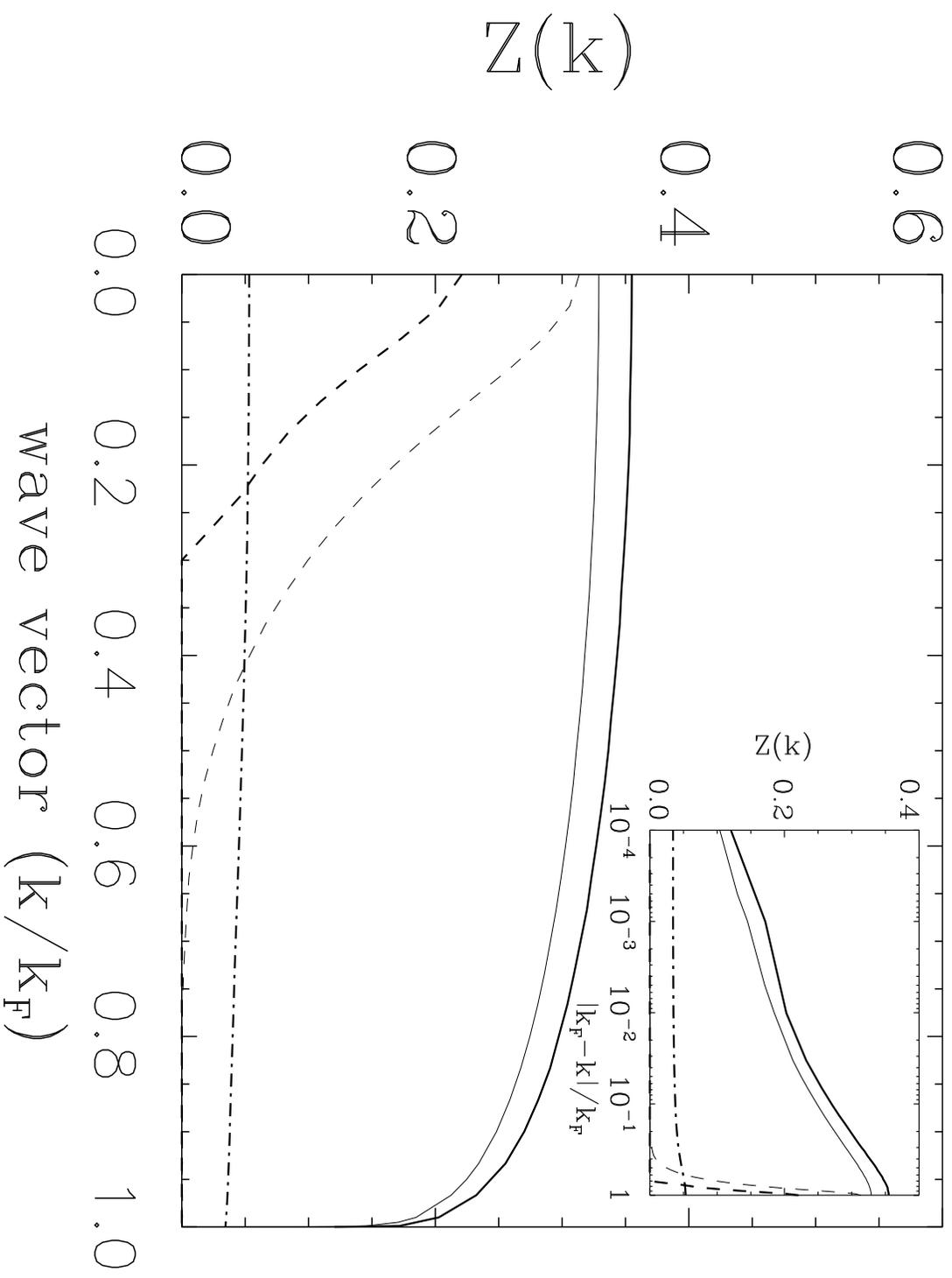
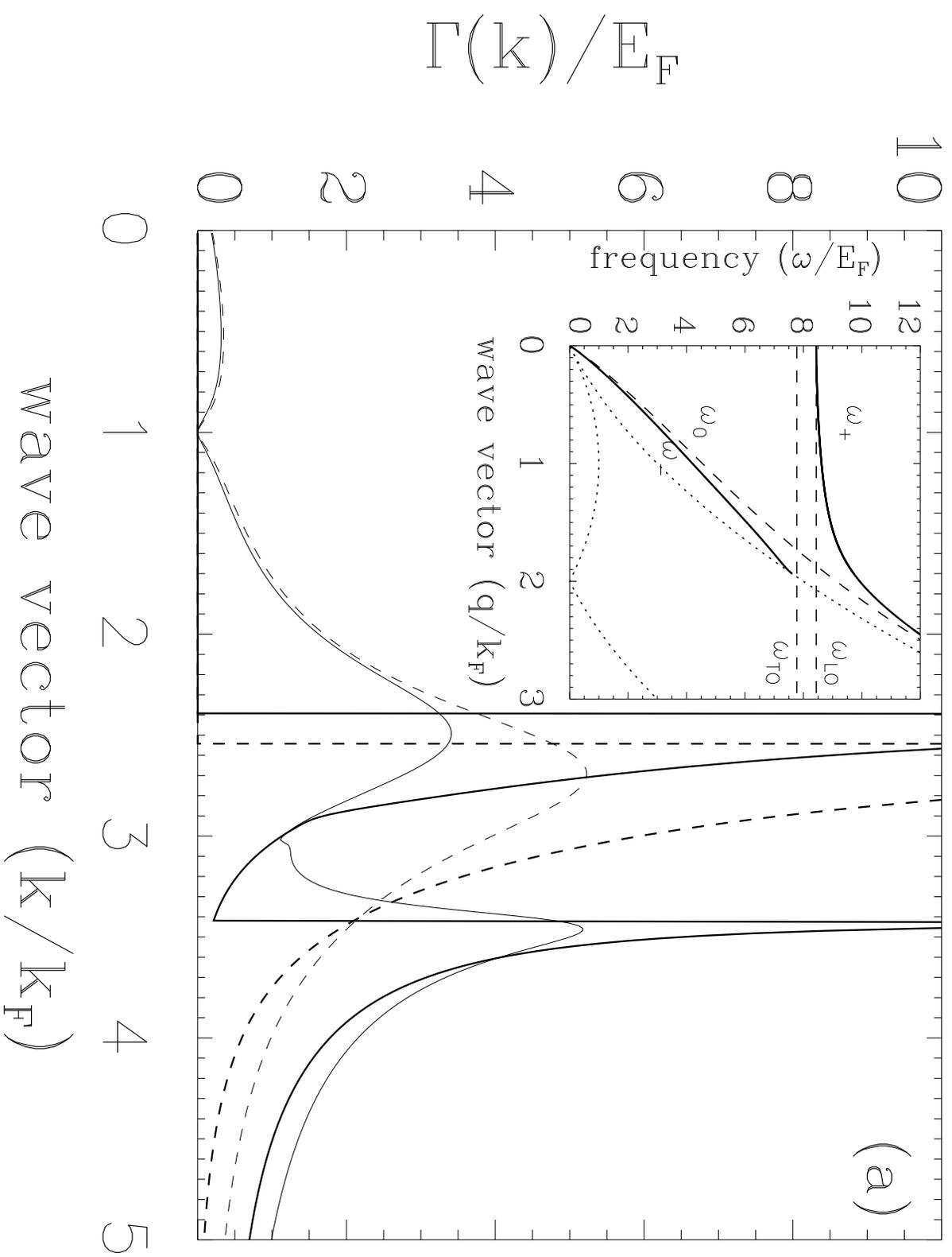
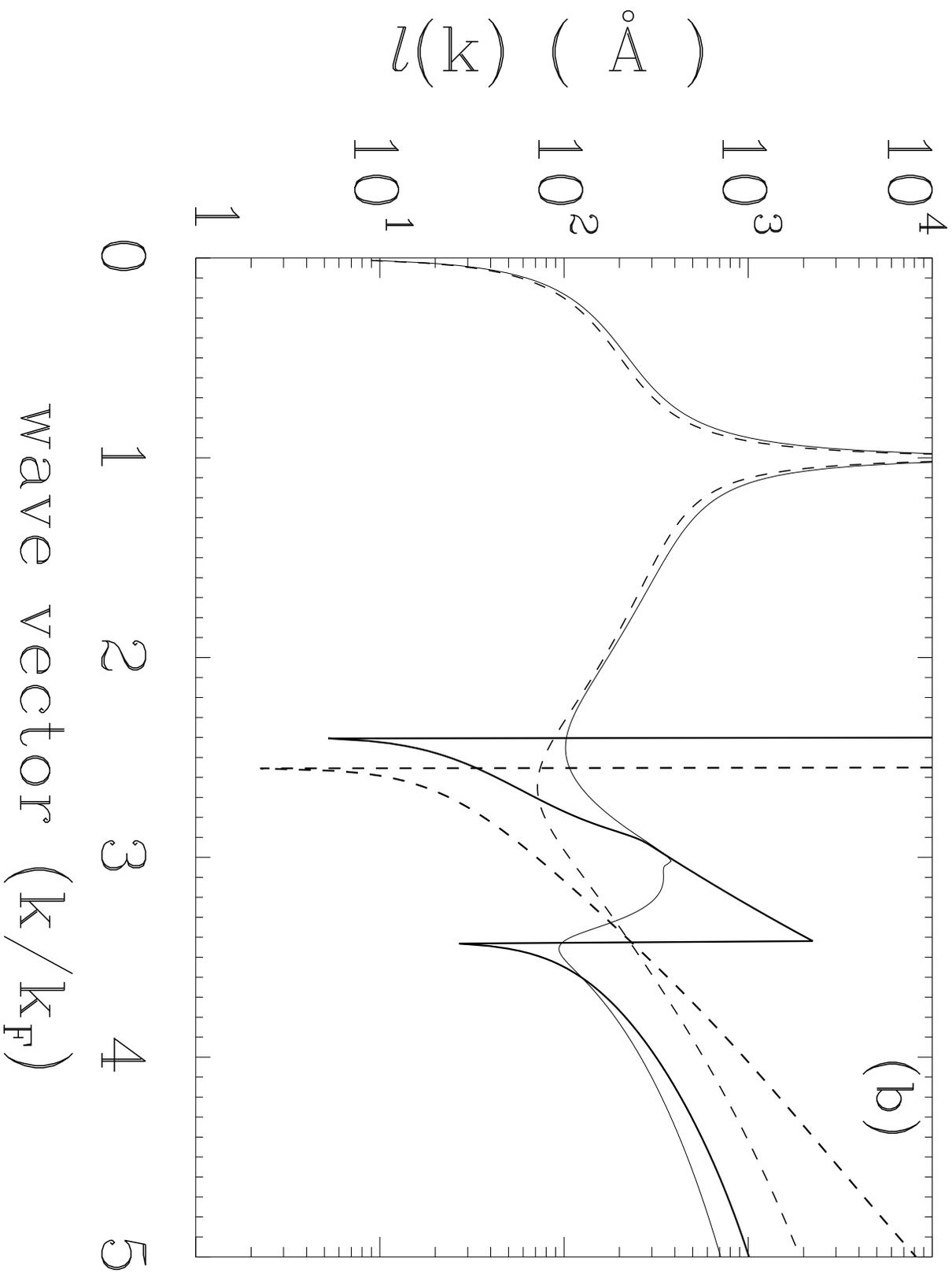
\end{document}